\documentclass[10pt,aps,twocolumn,superscriptaddress,preprintnumbers,nofootinbib]{revtex4-1} 
\usepackage{amsmath,amssymb,graphicx,natbib,bm}
\usepackage{enumerate}
\usepackage{hyperref}
\usepackage{color}

\usepackage{empheq}
\usepackage{slashed}

\newcommand{\beq}{\begin {equation}}  
\newcommand{\eeq}{\end   {equation}} 
\newcommand{\bea}{\begin {eqnarray}} 
\newcommand{\eea}{\end   {eqnarray}}  
\newcommand{\baa}{\begin {array}   } 
\newcommand{\eaa}{\end   {array}   }     
\newcommand{\bit}{\begin {itemize} }
\newcommand{\eit}{\end   {itemize} }
\newcommand{\be }{\begin {equation}} 
\newcommand{\ee }{\end   {equation}}
\newcommand{\nn }{\nonumber        }

\begin{document}


\preprint{ACFI-T16-35}


\title{Twin Higgs With $\mathbb{T}$-parity}

\author{Jiang-Hao Yu}
\email{jhyu@physics.umass.edu}
\affiliation{Amherst Center for Fundamental Interactions, Department of Physics, University of Massachusetts-Amherst, Amherst, MA 01003, U.S.A.}


\begin{abstract}

In twin Higgs models, a discrete $\mathbb{Z}_2$ symmetry between the standard model Higgs and the twin Higgs is introduced
to address the hierarchy problem. 
In this work, we propose another discrete symmetry in twin Higgs: the $\mathbb{T}$ parity, 
which maps the twin Higgs quadruplet into its mirror copy. 
The $\mathbb{T}$ parity brings us a whole group of $\mathbb{T}$-odd particles, 
and leads to a promising dark matter candidate.  
We present one realization of the $\mathbb{T}$ parity twin Higgs scenario by implementing the $SU(2)_L \times SU(2)_R \times U(1)_1 \times U(1)_2$ gauge symmetry in twin Higgs model. 
In this specific setup, the $\mathbb{T}$-odd $U(1)$ gauge boson could be the dark matter candidate, and the $\mathbb{T}$-odd particles
have very distinct and interesting phenomenology.
\end{abstract}

\maketitle


\section{Introduction}
\label{sec:intro}

In the standard model (SM), the large, quadratically divergent radiative corrections to the Higgs mass parameter destabilize the electroweak scale,
which is known as hierarchy problem.
The dynamical solution to this problem is to introduce a new symmetry which protects the Higgs mass against large radiative corrections.
Under this direction are weak scale supersymmetry, composite Higgs and twin Higgs, etc.  
Recently the twin Higgs scenario~\cite{Chacko:2005pe} attracts lots of attentions. 
And there have been studies on possible ultraviolet completion of the model~\cite{Craig:2014aea, Beauchesne:2015lva, Harnik:2016koz, Yu:2016bku} and on the twin particle phenomenology~\cite{Burdman:2014zta, Craig:2015pha, Cheng:2015buv, Chacko:2015fbc}. 
%
%
In the twin Higgs models, a twin Higgs doublet is introduced, and is mapped to the SM Higgs doublet through a discrete $Z_2$ symmetry. 
%
The $Z_2$ symmetry induces an accidental $U(4)$ global symmetry in the Higgs sector, which ensures that the SM Higgs boson becomes pseudo-Goldstone boson of the global symmetry breaking. 
Therefore, the $Z_2$ symmetry in twin Higgs protects the Higgs mass against large radiative corrections.

The twin Higgs models not only address the hierarchy problem, but also provide 
rich cosmological implications~\cite{Craig:2015xla, Garcia:2015loa, Freytsis:2016dgf, Prilepina:2016rlq, Chacko:2016hvu, Craig:2016lyx}, such as dark matter candidate, dark radiation, etc. 
In the mirror twin Higgs model~\cite{Chacko:2005pe, Craig:2014aea, Craig:2015pha}, the twin sector is only charged under new mirror SM gauge group and is connected to the SM sector via the $\mathbb{Z}_2$ symmetry. Thus the twin particles are completely neutral
under the SM gauge symmetry, and  could only talk to the SM sector through the Higgs boson. 
The twin particles in the mirror sector belong to a hidden dark sector, and it is the twin gauge symmetry
that stabilizes the dark matter candidate.
Typically the dark matter candidate could be twin-neutrino, twin-onium, etc~\cite{Craig:2015xla, Freytsis:2016dgf, Chacko:2016hvu, Craig:2016lyx}.
%
It provides us very interesting  cosmological consequence. 

In this work, we propose a twin Higgs scenario in which a $\mathbb{T}$ parity
is introduced to stablize the hidden dark sector in the twin Higgs models. 
Unlike the mirror gauge symmetry in mirror twin Higgs, which 
leads to stable dark matter candidate,
the $\mathbb{T}$ parity is a discrete symmetry. 
%
%
The $\mathbb{T}$ parity has been introduced and investigated in the little Higgs model~\cite{Cheng:2003ju, Cheng:2004yc}, to 
avoid the tight constraints from electroweak precision  tests and to 
introduce dark matter candidate in little Higgs. 
In our setup, two $U(4)$ twin Higgs quadruplets $H_1$ and $H_2$ are introduced, and 
an exchange symmetry in the Higgs sector is imposed:
\bea
	\mathbb{T}\,\, {\rm parity}: \,\, H_1 = \left(\begin{array}{c} 
	H_{1\rm SM} \\ H_{1\rm twin} \end{array}\right) \leftrightarrow H_2 = \left(\begin{array}{c} 
	H_{2\rm SM} \\ H_{2\rm twin} \end{array}\right),
\eea
The exchange symmetry between $H_1$ and $H_2$ is identified as the $\mathbb{T}$-parity.
Under the $\mathbb{T}$-parity, one combination of the Higgs quadruplet 
is $\mathbb{T}$-odd, while another is $\mathbb{T}$-even.
Not only the Higgs sector is doubled, 
the gauge and fermion structure could also be doubled through the $\mathbb{T}$-parity. 
Thus the $\mathbb{T}$-parity introduces a whole group of hidden particles, 
such as  $\mathbb{T}$-odd Higgs, $\mathbb{T}$-odd top partner, $\mathbb{T}$-odd gauge boson, etc.
This leads to a different cosmological implications than the ones in the mirror twin Higgs. 
%
To be specific, we present a realization of the $\mathbb{T}$ parity twin Higgs scenario by implementing the $SU(2)_L \times SU(2)_R \times U(1)_1 \times U(1)_2$ gauge symmetry in twin Higgs model. 
Effectively, this model can be viewed as the $\mathbb{T}$-parity extension of the left-right twin Higgs model~\cite{Chacko:2005un}. 
In this setup, the $\mathbb{T}$-odd particles are the top partner $T'$, the $\mathbb{T}$-odd Higgs $H'$, and the $\mathbb{T}$-odd $U(1)$ gauge boson $B'$, which is identified as the dark matter candidate. 
This $\mathbb{T}$-parity realization could be extended to other twin Higgs scenarios.

The paper is organized as follows. 
In Section II we introduce the twin Higgs model with $\mathbb{T}$-parity, and 
then write down the Lagrangian for $\mathbb{T}$-even and -odd particles in Section III. 
Section IV discusses the twin Higgs mechanism. 
In Section V and VI we investigate the model constraints on the  $\mathbb{T}$-even and -odd particles, 
respectively. 
Finally we conclude in Section VII.


\section{Twin Higgs Model with T-parity}
\label{sec:leftright}

We consider a two twin Higgs scenario, in which two $U(4)$ invariant Higgs quadruplets are introduced:
\bea
	H_1 \equiv \left(\begin{array}{c} H_{1L} \\ H_{1R} \end{array}\right),\qquad H_2 \equiv \left(\begin{array}{c} H_{2L} \\ H_{2R} \end{array}\right).
\eea
%
The tree-level Higgs potential preserves an approximate global symmetry $U(4)_1 \times U(4)_2$:
\bea
	V(H_1, H_2) = -\mu^2 (|H_1|^2 + |H_2|^2) + 
	\lambda \left[|H_1|^4 + |H_2|^4\right]. 
	\label{eq:Vtree}
\eea
There are two discrete symmetries in the scalar potential: 
\bit
\item the $\mathbb{Z}_2$ symmetry between $H_L$ and $H_R$: it maps the twin Higgses into the SM Higgses: $H_{1R} \xrightarrow{\mathbb{Z}_2} H_{1L}$, $H_{2R} \xrightarrow{\mathbb{Z}_2} H_{2L}$;
\item the $\mathbb{T}$-parity between $H_1$ and $H_2$: it maps the two Higgs quadruplets into each other: $H_{1} \xrightarrow{\mathbb{T}} H_{2}$.
\eit
%
The tree-level potential in Eq.~\ref{eq:Vtree} is invariant under both the $\mathbb{Z}_2$ symmetry and the T-parity.

\begin{figure}
  \includegraphics[width=0.36\textwidth]{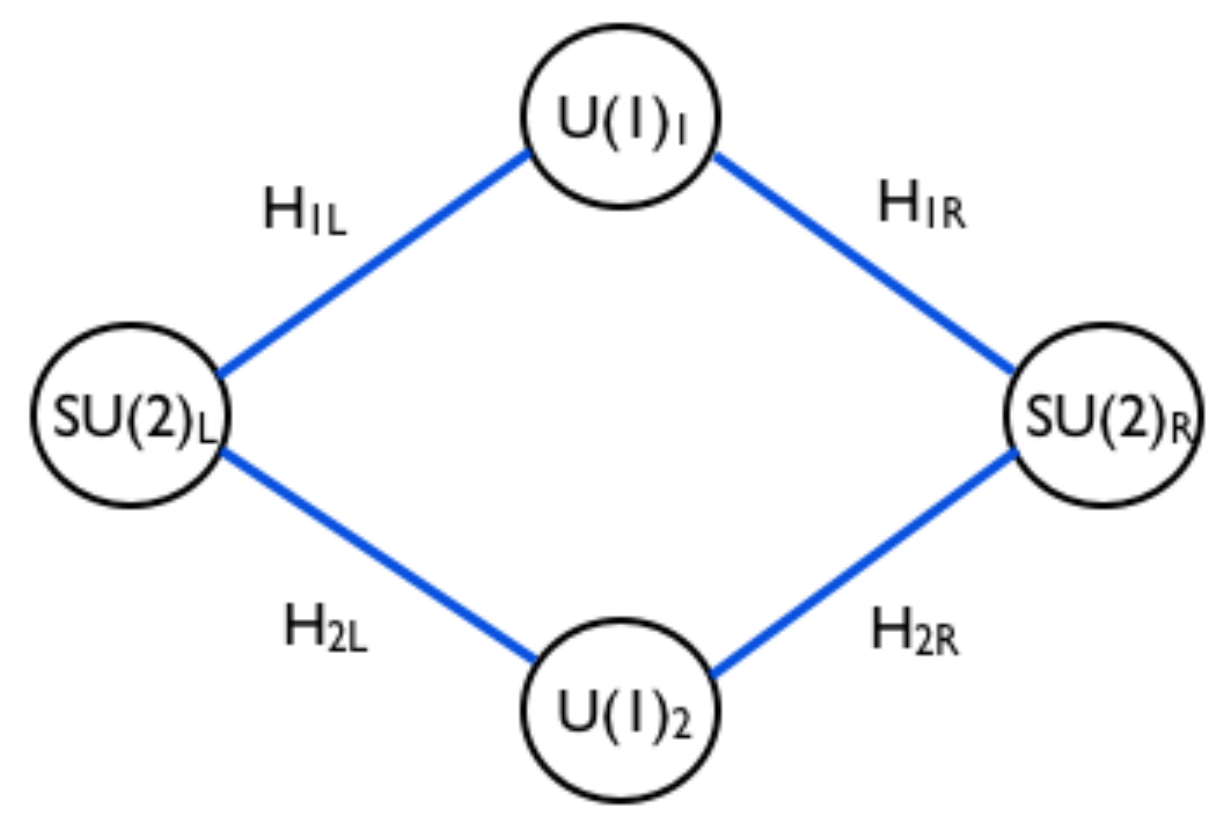}
\caption{\label{fig:moose} The "Moose notation"~\cite{Georgi:1985hf} diagram of this model: the gauged symmetry $SU(2)_L \times SU(2)_R \times U(1)_1 \times U(1)_2$
is represented by the solid circle, and the Higgs quadruplets $H_1$ and $H_2$ are represented by the links in between. 
The $\mathbb{Z}_2$ symmetry acts as: $H_{1L} \leftrightarrow H_{1R}$, $H_{2L} \leftrightarrow H_{2R}$, $SU(2)_L \leftrightarrow SU(2)_R$. And the $\mathbb{T}$-parity acts as: $H_{1L} \leftrightarrow H_{2L}$, $H_{1R} \leftrightarrow H_{2R}$ and $U(1)_1 \leftrightarrow U(1)_2$. }
\end{figure}

Due to the negative mass squared of $H_1$ and $H_2$, both $H_1$ and $H_2$ develop equal vacuum expectation values (VEVs): 
\bea
\langle H_1 \rangle = \langle H_2 \rangle = 
\left(\begin{array}{c} 0 \\ 0\\ 0\\ \mu/\sqrt{2\lambda}  \end{array}\right) = 
\left(\begin{array}{c} 0 \\ 0\\ 0\\ f/\sqrt2 \end{array}\right),
\label{eq:VEVs}
\eea 
which 
breaks $U(4)_{V} \to U(3)_{V}$ and leaves $U(4)_{A}$ unbroken.  
Here $U(4)_V$ is the diagonal subgroup of $U(4)_1 \times U(4)_2$, and $U(4)_{A} $ is the coset group.
Therefore, the $\mathbb{T}$-parity is still exact after the global symmetry breaking $U(4)_{V} \to U(3)_{V}$. 
%
%
The $U(4)_V$ is explicitly broken by gauging an $SU(2)_L \times SU(2)_R \times U(1)_1 \times U(1)_2$ subgroup.
Here $H_{1L}$ and $H_{2L}$ are doublets under $SU(2)_L$ gauge symmetry, while $H_{1R}$ and $H_{2R}$ are doublets under $SU(2)_R$ gauge symmetry.
$H_1$ is only charged under $U(1)_1$ gauge group, while $H_2$ is charged under $U(1)_2$.
In terms of the  "Moose notation"~\cite{Georgi:1985hf}, we exhibit the Moose and linked fields in Fig.~\ref{fig:moose}. 
Thus the two $U(4)$ invariant Higgs fields have the following gauged Lagrangian:
\bea
{\mathcal L} = (D_\mu H_1)^\dagger D^\mu H_1 
+ (D_\mu H_2)^\dagger D^\mu H_2 - V(H_1, H_2).
\eea
%
%
The covariant derivatives are
\bea
	D_\mu H_1 &=& \partial_\mu H_1 + i g {\mathbf W}_\mu H_1 + i g' {\mathbf B}_{1\mu} H_1,\nn\\
	D_\mu H_2 &=& \partial_\mu H_2 + i g {\mathbf W}_\mu H_2 + i g' {\mathbf B}_{2\mu} H_2,
\eea
where 
\bea
	{\mathbf W}_\mu = \frac{1}{2}\left(\begin{array}{cc} W^a_{L\mu} \tau^a & 0 \\ 0 &  W^a_{R\mu} \tau^a  \end{array}\right), 
	\,
	{\mathbf B}_{i\mu} = \frac{1}{2}\left(\begin{array}{cc} B_{i\mu}  & 0 \\ 0 &  B_{i\mu} \end{array} \right).
\eea
The gauged Lagrangian is invariant under the $\mathbb{Z}_2$ mapping $SU(2)_L \leftrightarrow SU(2)_R$, and the $\mathbb{T}$-parity mapping: $U(1)_1 \leftrightarrow U(1)_2$.

{\renewcommand{\arraystretch}{1.5} 
\begin{table}[ht]
\begin{center}
\begin{tabular}{c|  c c  c c}
                     & $SU(2)_L$ & $U(1)_1$ & $U(1)_2$ & $SU(2)_R$ \\ 
\hline 
$H_{1L}$               &  {\bf 2}          &$ \frac{1}{2}$        & $ 0 $ &  {\bf 1 }  \\
$H_{1R}$               &  {\bf 1}          &$ \frac{1}{2}$        & $ 0 $ &  {\bf 2 }  \\
$H_{2L}$               &  {\bf 2}          &$ 0 $        & $ \frac{1}{2}$ &  {\bf 1 }  \\
$H_{2R}$               &  {\bf 1}          &$ 0 $        & $ \frac{1}{2}$ &  {\bf 2 }  \\  
\hline                                                                        
$q_L$                &  {\bf 2}          &$ \frac{1}{6}$        & $ \frac{1}{6}$ &  {\bf 1 }  \\
$q_R$               &  {\bf 1}          &$ \frac{1}{6}$        & $ \frac{1}{6}$ &  {\bf 2 }  \\
$\ell_L$                &  {\bf 2}          &$-1$        & $ -1$ &  {\bf 1 }  \\
$\ell_R$               &  {\bf 1}          &$ -1$        & $ -1$ &  {\bf 2 }  \\
$T_{1L,R}$           &  {\bf 1}          &$+\frac{2}{3}$        & $0$ &  {\bf 1 }  \\
$T_{2L,R}$           &  {\bf 1}          &$0$        & $+\frac{2}{3}$ &  {\bf 1 }  \\
\hline
\end{tabular}
\end{center}
\caption{
The particle contents and their quantum numbers in the model. Here 
$T_{1}$ and $T_{2}$ are new vector-like top singlets. 
}
\label{tab:quantum}
\end{table}

To generate the light Higgs boson mass without quadratic divergence, 
we introduce two vector-like top singlets: $T_{1}$ and $T_{2}$. 
They are mapped into each other under $\mathbb{T}$-parity: $T_1 \leftrightarrow T_2$.
Adapting the matter contents in the left-right twin Higgs~\cite{Chacko:2005un}, we have the SM fermion contents 
(for simplicity, we only write down the third generation fermions) and the new fermions as follows: 
\bea
q_L &=& \left(\begin{array}{c} t_L \\ b_L  \end{array}\right), 
\ell_L = \left(\begin{array}{c} \nu_L \\ \tau_L  \end{array}\right), T_{1L}, T_{2L},\nn\\
q_R &=& \left(\begin{array}{c} t_R \\ b_R  \end{array}\right)
\ell_R = \left(\begin{array}{c} \nu_R \\ \tau_R  \end{array}\right), T_{1R}, T_{2R}.
\eea
Their quantum number assignments are listed in Table~\ref{tab:quantum}. 
The kinetic terms of the fermion Lagrangian are
\bea
 {\mathcal L}_{\rm ferm} &=& \overline{q_{L,R}} \gamma^\mu D_\mu q_{L,R} 
 + \overline{\ell_{L,R}} \gamma^\mu D_\mu \ell_{L,R} 
\nonumber\\
  &+& \overline{T_{1L,R}} \gamma^\mu D_\mu T_{1L,R}  + \overline{T_{2L,R}} \gamma^\mu D_\mu T_{2L,R},
\eea
where 
\bea
	D^\mu T_{1,2} &=& \partial^\mu T_i +  ig' Y B_{1,2}^\mu T_{1,2}, \\
	D^\mu q_{L,R} &=& \partial^\mu q_{L,R} +  \frac{ig}{2} W_{L,R}^{\mu a}  \tau^a  q_{L,R} \\\nonumber
	&&+  i g' Y (B_1^\mu + B_2^\mu) q_{L,R}, \nn\\
	D^\mu \ell_{L,R} &=& \partial^\mu  \ell_{L,R} +  \frac{ig}{2} W_{L,R}^{\mu a}  \tau^a   \ell_{L,R} \\\nonumber
	&&+  i g' Y (B_1^\mu + B_2^\mu)  \ell_{L,R}.
\eea
%
The top quark sector contains the SM top quark and new vectorlike tops.  
The top Yukawa Lagrangian is
\bea
 -{\mathcal L}_{\rm top} &=&  y_{1L} \overline{Q_L} H_{1L}  T_{1R} +  y_{1R} \overline{Q_R} H_{1R}  T_{1R} 
 + M \overline{T_{1L}} T_{1R} \\
 &+& y_{2L} \overline{Q_L} H_{2L}  T_{2R} +  y_{2R} \overline{Q_R} H_{2R} T_{2R} 
  + M \overline{T_{2L}} T_{2R} + h.c..\nn
\eea
Due to the $\mathbb{Z}_2$ symmetry and the $\mathbb{T}$-parity,
we have $y_{1L} =y_{2L} =y_{1R}  = y_{2R} = y$.
We obtain the top Yukawa coupling and top quark mass from the top Yukawa Lagrangian.  
Without introducing any more extra matter fields, all other SM quarks and leptons
can get their masses from the
 non-renormalizable terms
\bea
	-{\mathcal L}_{\rm Yuk} 
	&=& y_d \frac{\overline{q_L} H_{1L} H_{1R}^\dagger q_R + \overline{q_L} H_{2L} H_{2R}^\dagger q_R}{\Lambda} \nn\\
	&+&  y_\ell \frac{ \overline{\ell_L} H_{1L} H_{1R}^\dagger \ell_R
	+ \overline{\ell_L} H_{2L} H_{2R}^\dagger \ell_R }{\Lambda} \nn\\
	&+& y_u \frac{\overline{q_R} H_{1R}^\dagger H_{1L} q_L + \overline{q_R} H_{2R}^\dagger H_{2L} q_L  }{\Lambda}  + h.c.
\eea
Once the field $H_{iR}$ acquires a VEV of order $f$, the nonrenormalizable 
Lagrangian generates effective Yukawa couplings for the light quarks and leptons with the order of $f/\Lambda$, which 
is the typical size of the familiar Yukawa couplings in the SM~\cite{Chacko:2005un}.
In addition, we can write down the term 
\bea
	\frac{ \overline{\ell^C_R} H_{1R} H_{1R}^C \ell_R + \overline{\ell^C_R} H_{2R} H_{2R}^C \ell_R}{\Lambda}
\eea
which generates large Majorana masses for the $\nu_R$: $f^2/\Lambda$. 
Thus the small neutrino masses could be obtained via the seesaw mechanism.


\section{$\mathbb{T}$-Even/Odd Lagrangian}

The $\mathbb{T}$-parity is an exact symmetry of the Lagrangian. 
We could redefine the fields in the Lagrangian to have all the fields in the Lagrangian 
to be either $\mathbb{T}$-parity even or odd. 
We note that the fields $W_{L,R}^\mu$, $q_{L,R}$ are $\mathbb{T}$ even,
but $H_{1,2}, B_{1,2}^\mu, T_{1,2}$ are undetermined. 
Thus we define the following combinations:
%
\bea
	H &=& \frac1{\sqrt2}\left(H_{1}  + H_{2} \right), B^\mu = \frac1{\sqrt2}\left(B_1^\mu  + B_2^\mu \right), \nn\\
	H' &=& \frac1{\sqrt2}\left(H_{1}  - H_{2} \right), B'^\mu = \frac1{\sqrt2}\left(B_1^\mu  - B_2^\mu \right),\nn\\
	T_{L,R} &=& \frac1{\sqrt2}\left(T_{1L,R}  + T_{2L,R}  \right), \nn\\
	T'_{L,R} &=& \frac1{\sqrt2}\left(T_{1L,R}  - T_{2L,R} \right).
	\label{eq:Tevenodd}
\eea
Under these redefinition, we have 
\bit
\item $\mathbb{T}$-parity even fields: $H, B^\mu, T_{L,R}$, and $W_{L,R}^\mu$, $q_{L,R}$;
\item  $\mathbb{T}$-parity odd fields: $H', B'^\mu, T'$ with 
$H' \leftrightarrow -H', \quad B' \leftrightarrow -B', \quad T' \leftrightarrow - T'$.
\eit
Since the $\mathbb{T}$-parity is exact, it could be served as the origin 
of the dark matter symmetry, which could stabilize the dark matter candidate. 
Therefore, this model naturally explain the origin of the dark matter. 

The two $U(4)$ invariant quadruplets become
\bea
	H \equiv \left(\begin{array}{c} H_{L} \\ H_{R} \end{array}\right),\qquad H' \equiv \left(\begin{array}{c} H'_{L} \\ H'_{2} \end{array}\right).
\eea
According to Eq.~\ref{eq:VEVs} and the $\mathbb{T}$-parity, the VEVs of the $H$ and $H'$ are 
\bea
\langle H \rangle &=& \frac{1}{\sqrt2}\left( \langle H_1 \rangle + \langle H_2 \rangle \right) = 
\left(\begin{array}{c} 0 \\ 0\\ 0\\ f  \end{array}\right), \nn\\
 \langle H'\rangle &=& \frac{1}{\sqrt2}\left( \langle H_1 \rangle - \langle H_2 \rangle \right) =0.
\eea
The $\mathbb{T}$-odd field $H'$ has no VEV.
Therefore, the global symmetry breaking is $U(4) \to U(3)$ while $U(4)'$ is unbroken.
This can also be seen from the scalar  potential for $H$ and $H'$:
\bea
	V(H, H') &=& -\mu^2 (|H|^2 + |H'|^2) \\ 
	&+&\lambda \left[(|H|^2 + |H'|^2)^2 + (H^\dagger H' + H'^\dagger H)^2\right].\nn
	\label{eq:Vtree2}
\eea 
The deepest minima of the potential exist at either $(\langle H \rangle, \langle H'\rangle) = (f , 0)$ or 
$(\langle H \rangle, \langle H'\rangle) = (0, f)$.

The symmetry breaking pattern is
\bea
  \textrm{global   symmetry:} && \quad U(4)  \to U(3) , \nn\\
  \textrm{gauge    symmetry:}  && \quad SU(2)_L  \times SU(2)_R \times U(1) \times U(1)' \nn\\
  &&\quad \to SU(2)_L \times U(1)_Y \times U(1)'.\nn
\eea 
Let us parametrize the fields $H$ nonlinearly in terms of the nonlinear sigma field
\bea
	H = \exp\left[\frac{i}{f}\left(\begin{array}{c|cc}
		\bm{0}_{2\times 2}&\bm{0}_{1 \times 2}&\bm{h}\\\hline
		\bm{0}_{2 \times 1}&0&C\\ 
		\bm{h}^{\ast}&C^{\ast}&N
		\end{array} \right)\right]\left(\begin{array}{c} \bm{0}_{1 \times 2} \\\hline 0 \\ f \end{array}\right),
\eea
where the field $\bm{h}$ denotes the SM Higgs doublet $\bm{h} = \left(\begin{array}{c}  h^+  \\  h^0  \end{array}\right)$, 
and $C^\pm$ and $N$ are Goldstone bosons, which are absorbed by the $SU(2)_R \times U(1)$ gauge bosons.
Taking the expansion, the field $H$ takes the form
\bea
H=
\left(\begin{array}{c}
f\frac{i \bm{h}}{\sqrt{\bm{h}^{\dag}\bm{h}}}\sin\left( \frac{\sqrt{\bm{h}^{\dag}\bm{h}}}{f} \right)\\
0\\
f \cos\left( \frac{\sqrt{\bm{h}^{\dag}\bm{h}}}{f} \right)
\end{array} \right)
\simeq 
\left(\begin{array}{c}
i \bm{h}\\
0\\
f - \frac{1}{2  f}\bm{h}^{\dag}\bm{h}
\end{array} \right).
\eea
Here the field $H$ plays the role of  the twin Higgs as the original twin Higgs model. 
Another field $H'$ does not obtain VEV, and thus it is just another scalar quadruplet  in this model.

Using the fields in Eq.~\ref{eq:Tevenodd}, the kinetic Lagrangian in the scalar sector becomes
\bea
	{\mathcal L} &=& (D_\mu H^\dagger - ig' Y  B'_\mu H'^\dagger) (D^\mu H +  ig' Y  B'_\mu H') \nn\\
	&&+ (D_\mu H'^\dagger - ig' Y  B'_\mu H^\dagger) (D^\mu H' +  ig' Y  B'_\mu H),
\eea
where the covariant derivative is defined as
\bea
	D^\mu H  &=& \partial^\mu H + i g {\mathbf W}^\mu H + i g' {\mathbf B}^\mu H , \nn\\
	D^\mu H' &=& \partial^\mu H' + i g {\mathbf W}^\mu H' + i g' {\mathbf B}^\mu H'. 
\eea
Note the Higgs mechanism for the $\mathbb{T}$-odd field $B'^\mu$  is quite different from the typical case:
\bit
\item In typical case, for example, the $\mathbb{T}$-even field $B^\mu$ absorbs the CP odd component of the T-even $H$ and obtains its mass from its VEV $\langle H \rangle$;
\item The terms $\partial^\mu H' B'_\mu H$ and $B'^\mu B'_\mu H^\dagger H$ indicate that the $\mathbb{T}$-odd field $B'^\mu$ absorbs the CP odd component of the T-odd $H'$ but obtains its mass from VEV of the $H$.
\eit

%
%
%
The kinetic Lagrangian in the fermion sector becomes
\bea
	{\mathcal L} &=&  \overline{T} i\gamma^\mu (\partial_\mu + i g' Y  B_\mu) T  + \overline{T'} i\gamma^\mu (\partial_\mu + i g' Y  B_\mu) T' \nn\\
	&& -  g'\overline{T} \gamma^\mu B'_\mu T' - g' \overline{T'} \gamma^\mu B'_\mu T.
\eea
And the Yukawa Lagrangian in the top quark sector becomes
\bea
	{\mathcal L}_{\rm Yuk} 
	&=& \overline{q_L} H_{L} T_{R} + \overline{q_R} H_{R} T_{L} + M T_L T_R \nn\\
	&+& \overline{q_L} H'_{L} T'_{R} + \overline{q_R} H'_{R} T'_{L} + M T'_L T'_R+ h.c.	
\eea
We also obtain the Yukawa Lagrangian for the SM quarks and leptons:
\bea
	-{\mathcal L}_{\rm Yuk} 
	&=&  \frac{y_d \overline{q_L} H_{L} H_{R}^\dagger q_R + y_\ell \overline{\ell_L} H_{L} H_{R}^\dagger \ell_R }{\Lambda} \nn\\
	&+& y_u \frac{\overline{q_R} H_{R}^\dagger H_{L} q_L }{\Lambda}  + h.c.
\eea

\section{Twin Higgs Mechanism}

The gauge and Yukawa interactions break the global symmetry $U(4)$ explicitly, 
generate masses for the massless Higgs boson, and 
trigger the electroweak symmetry breaking.
We ultilize the Coleman-Weinberg (CW) potential to quantify the radiative corrections of the Higgs potential.
The one-loop CW potential in Landau gauge~\cite{Coleman:1973jx} is
\bea
	V_{\rm CW}(H) &=& \frac{1}{64\pi^2} {\rm STr}\left[ \Lambda^4 \left(\ln \Lambda^2 - \frac32\right) + 2 {\mathcal M}^2(H) \Lambda^2 \right. \nn\\
	&&\left. +  {\mathcal M}^4(H)\left(\ln\frac{{\mathcal M}^2(H)}{\Lambda^2} - \frac32\right)\right],
	\label{eq:Vcw}
\eea
where the super-trace STr is taken among all the dynamical fields that have the Higgs dependent masses.
The first term is the cosmological constant term, while the second term  is responsible for the quadratic divergence of the Higgs boson masses.
It is the third term that gives the scalar potential of the Higgs boson.

The Higgs dependent charged gauge boson masses are
\bea
	m_{W}^2 = \frac{1}{2} g_L^2 |H_L|^2, m_{W'}^2 = \frac{1}{2} g_R^2 |H_R|^2,
	\label{eq:Wmass}
\eea
where $g_L = g_R = g$ according to the $\mathbb{Z}_2$ symmetry.
The Higgs dependent neutral gauge boson masses are
\bea
	m_Z^2 &\simeq & \frac{1}{2} (g^2 + g_Y^2) |H_L|^2 - \frac{1}{2f^2} \frac{g_Y^4}{g^2} |H_L|^4,\nn\\
	m_{Z'}^2 &\simeq & \frac{1}{2} (g^2 + g'^2) f^2 - \frac{1}{2} (g^2 + g_Y^2) |H_L|^2 \nn\\
	&&  + \frac{1}{2f^2}  \frac{g_Y^4}{g^2} |H_L|^4.
\eea
The Higgs dependent top quark masses are
\bea
	m_t^2 &=& \frac{y^4 |H_L|^2 |H_R|^2}{(M^2 + y^2 f^2)},\nn\\
	m_T^2 &=& M^2 + y^2 f^2 - \frac{y^4 |H_L|^2 |H_R|^2}{(M^2 + y^2 f^2)}.	
\eea
For the $\mathbb{T}$-odd particles, we have 
\bea
	m_{B'}^2 &=& \frac{1}{2} g'^2 (|H_L|^2 + |H_R|^2) = \frac{1}{2} g'^2 Y^2 f^2,\nn\\
	m_{T'}^2 &=& M^2,
\eea
which have no dependence on the Higgs boson field, 
and thus are not relevant to the Higgs boson mass and potential.

Let us first discuss the quadratic dependence of the Higgs boson in the CW potential in Eq.~\ref{eq:Vcw}. 
Considering the contributions from the charged gauge bosons in Eq.~\ref{eq:Wmass},  
we have 
\bea
	V_{\rm CW} \supset 
	\frac{9\Lambda^2}{64\pi^2} (g_L^2 |H_L|^2 + g_R^2 |H_R|^2)
\eea
Only if the $\mathbb{Z}_2$  symmetry is imposed ($g_L = g_R = g$), 
there is no quadratic divergence on the Higgs boson mass from the charged gauge boson radiative corrections. 
%
%
Similarly, we obtain the CW potential from the neutral gauge bosons and the top quark sector
\bea
	V_{\rm CW} &\supset& \frac{9\Lambda^2}{64\pi^2} (m_{Z}^2 + m_{Z'}^2)  \simeq \Lambda^2 f^2, \\
	V_{\rm CW} &\supset& -\frac{3\Lambda^2}{8\pi^2} (m_{t}^2 + m_{T}^2)  \simeq \Lambda^2 f^2.
\eea
In summary, the quadratic part of the CW potential {\it accidentally} respect the original $U(4)$ symmetry due to the
$\mathbb{Z}_2$ symmetry, which is the twin Higgs mechanism.
Thus the Higgs mass does not receive any quadratic divergent contributions due to the twin Higgs mechanism. 

Now let us analyse the radiatively generated Higgs potential. The leading Higgs potential is parametrized as
\bea
	V(h^\dagger h) 
	= a \sin^2\left( \frac{\sqrt{\bm{h}^{\dag}\bm{h}}}{f} \right) + b\sin^4\left( \frac{\sqrt{\bm{h}^{\dag}\bm{h}}}{f} \right),
\eea
where $a$ and $b$ are coefficients calculated from the one-loop CW potential. 
%
%
The gauge boson contributions are
\bea
	a &=& \frac{3 }{64\pi^2}g^4 f^4\left(\log\frac{\Lambda^2}{g^2 f^2/2} + 1\right)\nn\\
	&&+ \frac{3 g^2 (g^2 + 2 g'^2)f^4}{128\pi^2}\left(\log\frac{\Lambda^2}{(g^2 + g'^2)f^2/2} + 1\right),\\
	b &=& - a
	+ \frac{3 }{256\pi^2}(g^2 + g_Y^2) f^4\left[  \log\frac{(g^2+g'^2)f^2/2}{m_Z^2} - \frac12 \right],\nn
\eea
where $m_Z^2 = \frac12 (g^2 + g_Y^2)f^2 \sin^2 x$.
Due to  positive value of $a$, the gauge boson contributions could not 
trigger electroweak symmetry breaking. 
The top quark contributions are
\bea
a &=& -\frac{3}{8\pi^2} y^4 f^4\left(\log\frac{\Lambda^2}{M^2 + y^2 f^2} + 1\right),\nn\\
b&=& -a + \frac{3y_t^4 f^4  }{16\pi^2}
\left[  \log\frac{M^2 + y^2 f^2}{m_t^2} - \frac12 \right],
\eea
where the top-Yukawa coupling is defined as $y_t = y\frac{y f}{\sqrt{M^2 + y^2 f^2}}$,
and $m_t^2 = y_t^2 f^2 \sin^2 x$. 
Therefore, the electroweak symmetry breaking is triggered by the 
top quark contributions.
The Higgs boson mass is calculated via
\bea
	m_{\rm Higgs}^2 \simeq - \frac{a}{b} f^2,
\eea
But here we have $a \sim b$, and thus we obtain 
the Higgs mass is around $f^2$, which is too heavy. 
We need to add soft mass term, which only break the $\mathbb{Z}_2$ symmetry softly
but the $\mathbb{T}$-parity is still exact. 
The soft $\mathbb{Z}_2$ breaking term reads
\bea
	V_{\rm soft} = m^2 (H_{1A}^\dagger H_{1A} + H_{2A}^\dagger H_{2A}).
	\label{eq:soft}
\eea
The Higgs mass is then
\bea
	m_{\rm Higgs}^2 \simeq  \frac{a - m^2 f^2}{b} f^2.
\eea
Since the Higgs mass is measured to be 125 GeV, we could determine the 
soft mass parameter $m^2$ once we know the new physics scale $f$.

The $\mathbb{T}$-odd Higgs quadruplet $H'$ also receives radiative corrections. 
Let us denote component fields in the $H'$ as 
\bea
	H' = \left(\begin{array}{c} {H'}_L^{+} \\ {H'}_L^{0} + i {A'}_L^{0}  \\ {H'}_R^{+} \\ {H'}_R^{0} + i {A'}_R^{0} \end{array}\right).
\eea
Due to the exact $\mathbb{T}$-parity, although the $H'$ does not mix with the $H$, 
they share the same potential, as shown in Eq.~\ref{eq:Vtree2}.
Therefore, at tree-level, after the $H$ obtains its VEV $\langle H_L \rangle = f \sin \frac{v}{f}, \langle H_R \rangle = f \cos \frac{v}{f}$, 
we obtain the tree-level masses
\bea
m_{{H'}_L^{0}}^2 &=& 2\lambda f^2 \sin^2 \frac{v}{f}, \\
m_{{H'}_R^{0}}^2 &=& 2\lambda f^2 \cos^2 \frac{v}{f},
\eea 
while other components are massless. 
On the other hand, since $H'$ has no VEV at all~\footnote{%
It is  different from the case that $H'$ obtains its VEV~\cite{Chacko:2005un, Goh:2006wj}. 
If the $H'$ has VEV $\langle H'_R \rangle = f'$, 
the CP-odd scalar in $H'_R$ will be massless due to a global residue $U(1)_R$ symmetry. }, 
all the components receive radiative corrections in additional to the tree-level masses:
\bea
	m_{\rm all\,\,H'\,\, components}^2 \simeq \frac{1}{16\pi^2} g^4 f^2 \log\frac{\Lambda}{f}. 
\eea
Similar to radiative corrections, adding soft mass terms will also lift the masses of the $H'$ component fields. 
Given the soft mass terms in Eq.~\ref{eq:soft}, all $H'_L$ component fields obtain additional mass corrections:
\bea
	m_{\rm all\,\,H'_L\,\, components}^2 \simeq m^2. 
\eea
If we add soft mass terms for the $H'_R$, all the $H'_R$ component fields will
also receive corrections from the soft term.
%
Therefore, the masses of the $H'$ components will be the sum of all kinds of mass corrections. 
Since the soft mass terms origin from the ultraviolet physics, 
the masses of the $H'$ component fields are quite sensitive to the UV completion of this model.

\section{Model Constraints}


The strongest experimental constraints on the model come
from direct searches at the LHC on the new $\mathbb{T}$-even particles: 
the new gauge bosons $W'$, $Z'$ and the colored heavy top $T$.
Both ATLAS and CMS investigated the exotic $W', Z', T'$ resonances. 
The up-to-dated bounds on the masses of these resonances are summarized as follow:
\bit
\item If the right-handed $W'$ decays to right-handed neutrino and lepton, 
the high mass resonance searches in the lepton plus transverse missing energy final states
put strong constraint on the $W'$ mass. 
In our setup, the right-handed neutrino masses are around the scale $f$, to 
have see-saw mechanism to generate the neutrino mass. 
Therefore, the right-handed $W'$ will dominantly decay to di-jet and single top final states. 
Due to huge QCD background in di-jet channel, we expect that the single top final states
provide us the tightest constraint on the $W'$ mass. 
Based on the 13 TeV CMS data with 12.9 fb$^{-1}$ luminosity~\cite{CMS:2016wqa}, the observed 95\% confidence level (CL) limits on the right-handed $W'$ is $m_{W'} > 2.6$ TeV~\footnote{%
The exclusion limit from ATLAS~\cite{Aad:2014xea} are slightly weaker because 8 TeV data was used in their analysis. Furthermore, flavor physics, such as the $K_S-K_L$ mixing, etc, also put constraint on the $W'$ mass, but it is weaker than the updated LHC constraints. }. 
But this is for the right-handed $W'$ with 100\% decay branching ratio to single top.
Recasting the updated exclusion limit, we obtain $m_{W'} > 1.67$ TeV. 

\item The dileptonic final states at the LHC put the strongest limit on the  $Z'$ gauge boson. 
Based on the 13 TeV ATLAS (13.9 fb$^{-1}$)~\cite{ATLAS:2016cyf} and CMS (2.9 fb$^{-1}$)~\cite{Khachatryan:2016zqb} data,
the observed 95\% confidence level (CL) limits on the sequential $Z'$ is $m_{Z'} > 4.05$ TeV. 
Taking into account the branching ratio of the dilepton (around 2.5\% in this model), we obtain the exclusion limit
$m_{Z'} > 2.56$ TeV.

\item The heavy top quark partner $T$ has been investigated at both the ATLAS~\cite{ATLAS:2016cuv} and CMS~\cite{Chatrchyan:2013uxa}.  The tightest constraint on $T$ come from the combination of the decay channels $T \to t Z$, $T \to b W$ and $T \to t h $.
Based on the 13 TeV ATLAS with 11.5 fb$^{-1}$ luminosity, the updated exclusion limit is around $850$ GeV after taking the branching ratios into account.
\eit
From above, we note that the tightest constraint comes from the $Z'$ dileptonic searches. 
Converting to constraints on the scale $f$, we obtain that the scale $f > 4.8$ TeV.
This introduces a tuning between the scale $f$ and the electroweak scale, 
which is the so-called little hierarchy problem. 
Note that mass bounds on $T$ is not so tight, because both the parameter $M$ and the scale $f$ 
contribute to the value of the $T$ mass. 
The vectorlike mass $M$ lifts the mass of $T$ and could keep the scale $f$ below 700 GeV. 
We could ultilize similar setup to lift the masses of the new gauge bosons while keep the scale $f$ below 1 TeV.  
Furthermore, lifting the new gauge boson masses will also address the indirect limits from electroweak precision tests. 
If the masses of the $W'$ and $Z'$ are not so heavy, the electroweak precision test put strong constraints 
on the model parameters~\cite{Hsieh:2010zr}. 
However, when the gauge boson masses are heavier than 2 TeV,  the electroweak precision constraints could be much weaeker~\footnote{Although the electroweak precision test might provide constraints on the mixing angle $s_L$ and $s_R$ between $t$ and $T$ for a sub-TeV $T$~\cite{Xiao:2014kba}, 
the $\mathbb{T}$-parity between $T$ and $T'$ could weaken the electroweak precision constraints.} 
 

There are ways to lift the new gauge boson masses by including new scalar fields which 
are charged under the gauge group of the model. 
To keep other sectors in this model unaffected, we need to assign such new scalar fields only 
play the role of giving the new gauge boson masses without interacting with other fermions or scalars. 
%
Here we suggest two ways to lift the gauge boson masses:
\bit
\item way I: introduce additional complex scalar $S$, which is only charged under $U(1)_1$ and $U(1)_2$. 
The same $U(1)$ charges are needed to keep the $\mathbb{T}$-parity exact. 
After this new scalar $S$ obtains its VEV $f'$, the $\mathbb{T}$-even gauge boson $Z'$ obtains its mass of order $\sqrt{f^2 + f'^2}$. 
In this way, the scale $f$ could escape the tight constraints from the dileptonic $Z'$ searches. 
We could lower the scale $f$ to be around 2 TeV in which the $W'$ mass bound plays the significant role. 

\item way II: introduce additional Higgs quadruplet $\tilde{H}$, which is charged under $SU(2)_L \times SU(2)_R$ and $U(1)_1 \times U(1)_2$. Similarly, the same $SU(2)$ and $U(1)$ charges are needed to keep the $\mathbb{T}$-parity exact.  
After the Higgs quadruplet $\tilde{H}$ obtains its VEV $f'$, both the  $W'$ and $Z'$ obtains their masses of order $\sqrt{f^2 + f'^2}$. 
In this way, we could lift the masses of the $W'$ and $Z'$ while keep the $f$ to be around 1 TeV. 
%
\eit
Therefore, although the current limits on the $W'$ and $Z'$ are tight, the scale $f$ could still be around TeV scale.
Furthermore, if we assume the new scalar masses are heavier than TeV scale by introducing soft mass terms, adding these new scalars will not affect the phenomenology below TeV scale. 
We will simply assume these new scalars are very heavy and only affects the $W'$ and $Z'$ masses. 

\begin{figure}[htp]
\begin{center}
\includegraphics[width=0.15\textwidth]{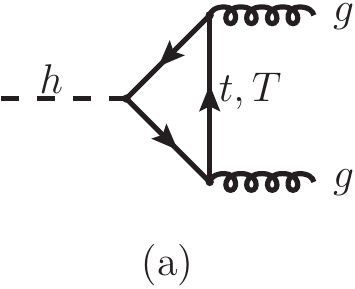} 
\includegraphics[width=0.15\textwidth]{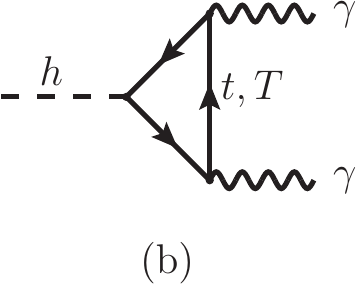} 
\includegraphics[width=0.15\textwidth]{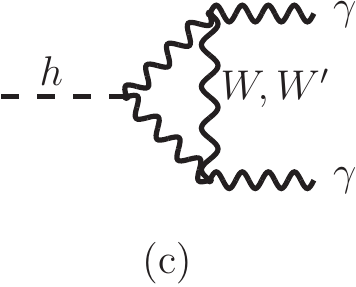} 
\caption{\small  The Feynman diagrams on the Higgs gluon gluon process (a) and Higgs to diphoton process (b,c). }
\label{fig:feynhiggs}
\end{center}
\end{figure}

\begin{figure}[htp]
\begin{center}
\includegraphics[width=0.35\textwidth]{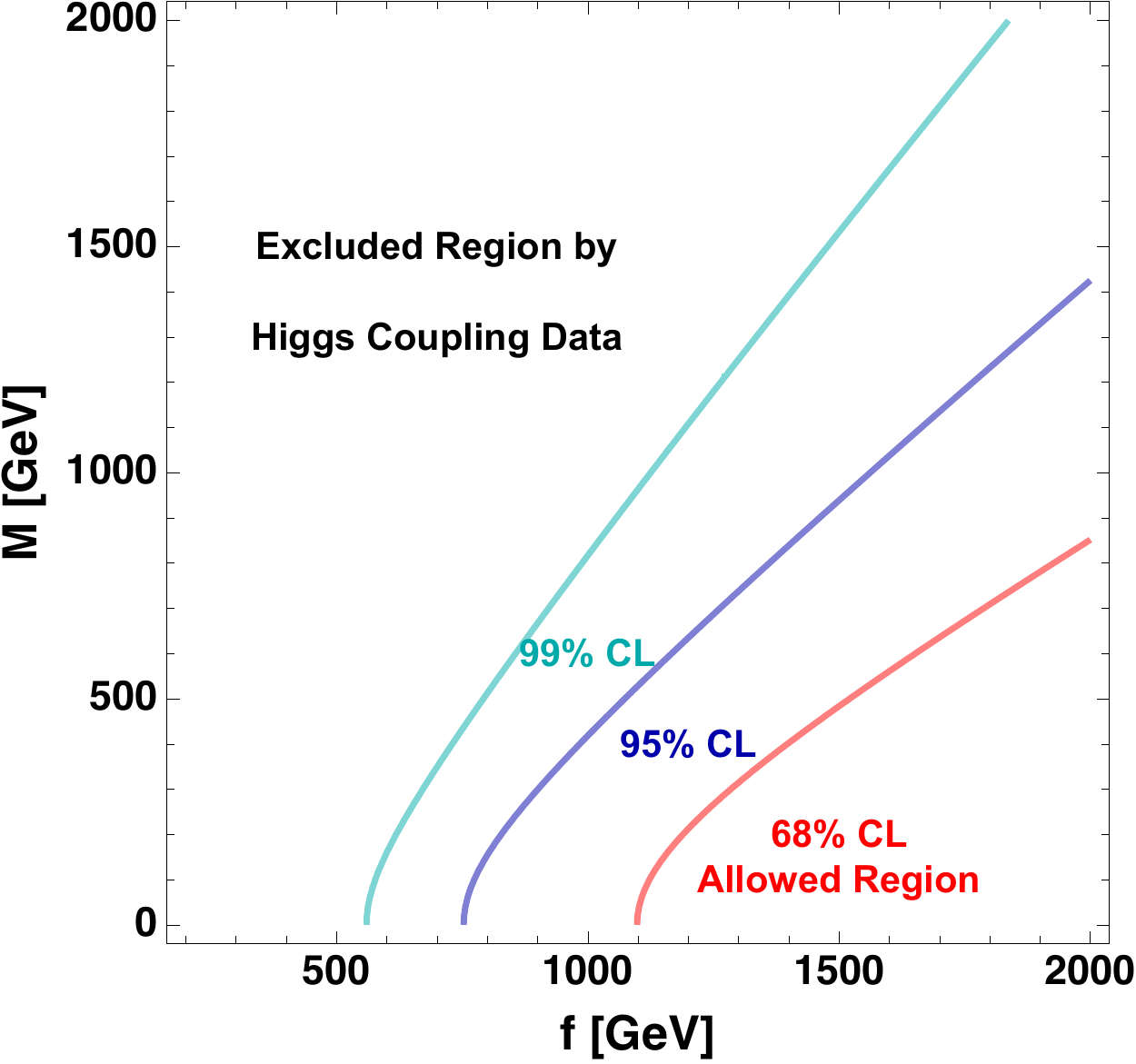} 
\caption{\small  The the allowed contours on $(f, M)$ at 68\%, 95\%, 99\% CLs, according to the  global fitting of the Higgs coupling data. }
\label{fig:higgs}
\end{center}
\end{figure}

Although adding new scalars could lower the scale $f$, 
the Higgs coupling measurements will constrain the scale $f$ whatever the new scalars exist or not.
%
%
Independent of the new scalar sector, the Higgs coupling measurements put strong limit on the model parameters $(f, M)$.   
The Goldstone boson nature of the Higgs boson modifies the Higgs couplings to the 
SM gauge boson and SM fermions by a factor $\left(1-\frac{v^2}{f^2}\right)$. 
Furthermore, due to the mixing between $t$ and $T$, the $htt$ and $hTT$ couplings are
further modified by the mixing angle. 
This will affect the Higgs couplings to the gluon fields via the loop effects, as  shown in Fig.~\ref{fig:feynhiggs}(a). 
Finally, due to the existence of the charged gauge bosons and charged inert scalars, 
the Higgs diphoton couplings are also modified, as shown in Fig.~\ref{fig:feynhiggs}(b,c). 
In principle, all the charged particles 
are involved in the diphoton loop. 
Given that the masses of the ${H'}^\mp_{L,R}$ depend on the soft mass terms, 
here for simplicity, we assume the ${H'}^\mp_{L,R}$ are heavy, and thus the
scalar contribution is negligible, compared to other contributions.

The relevant Higgs couplings are 
\bea
	&&hWW: \frac12 g^2 v\left(1-\frac{v^2}{f^2}\right), \quad hW'W': -\frac12 g^2 v \left(1-\frac{v^2}{f^2}\right), \nn\\
	&&htt: - \frac{y_t}{\sqrt2} c_{L}c_{R}, \quad hTT: - \frac{y}{\sqrt2}(s_{L}s_{R} -  c_{L}c_{R}\frac{v}{f}),
\eea
where $s_L$ and $s_R$ are mixing angle between $t$ and $T$, defined in the Appendix.  
Using these couplings, we calculate various Higgs signal strengths $\mu_{pp\to h_1 \to ii} =  \sigma(pp\to h_1) \textrm{Br}_{h_1 \to ii}/\sigma_{\rm SM}\textrm{Br}_{\rm SM}$.
Based on Higgs signal strengths at the 8 TeV LHC with 20.7 fb$^{-1}$ data~\cite{Aad:2015gba, Khachatryan:2014jba}, 
we perform a global fit on the model parameters. 
Fig.~\ref{fig:higgs} shows the allowed contours on $(f, M)$ at 68\%, 95\%, 99\% CLs. 
Depending on values of the mass parameter $M$, the scale $f$ could be as low as $750$ GeV at 95\% CL. 
As shown in the next section, the mass parameter $M$ determines the mass scale of the $\mathbb{T}$-odd top 
partner $T'$. 
Given the current limit on the $T'$ mass $m_{T'} > 0.9$ TeV, 
we determine the scale $f$ should be around 1.3 TeV at 95\% CL.


\section{$\mathbb{T}$-Odd Particle Phenomenology}

The signatures of the $\mathbb{T}$-odd particles provide  very distinct 
features from the original twin Higgs model. 
Due to the exact $\mathbb{T}$ parity, the $\mathbb{T}$-odd particles 
do not mix with the $\mathbb{T}$-even particles. 
Similar to the little Higgs models with $\mathbb{T}$ parity~\cite{Cheng:2003ju, Cheng:2004yc}, we assign the
$\mathbb{T}$-odd particles  belong to dark sector, and the lightest 
$\mathbb{T}$-odd particle (LTP) is the dark matter candidate.

In this left-right twin Higgs setup, 
%
The $\mathbb{T}$-odd particles are the dark gauge boson $B'$, the dark top $T'$ and all the component fields in $H'$. 
The $\mathbb{T}$-odd particle masses are $m_{T'} = M$, and $m_{B'} = \frac12 g' f$. 
As discussed above, the masses of the $H'$ strongly depend on the soft mass terms, and typically 
$m_{{H'}_{L,R}^{0}}^2 \sim \lambda f^2$, $m_{{H'}_{L,R}^{\pm}} \sim m_{{A'}_{L,R}^{0}} \sim m_{\rm soft}$. 
Depending on the soft mass term, the dark matter candidate could be
either ${A'}_{L,R}^{0}$ or the dark gauge boson $B'$. 
%
%
Here for simplicity, we assume the soft mass term, which is typically order of $y f$, is typically larger than $g f$, and 
take the dark gauge boson $B'$ as the dark matter candidate. 
In this case,  the dark matter signatures are quite different from the ones in the left-right twin Higgs model~\cite{Dolle:2007ce}.

Although the kinetic term of the dark gauge boson involves in $H$, there is no 
coupling between $B'$ and the Higgs boson, such as $B' B' h h$ and $B' B' h$ terms. 
The $B'$ typically interacts with the $\mathbb{T}$-odd $T'$ via the $\mathbb{T}$-even top quarks, 
and couples to the ${H'}^\pm_{L,R}$ and ${H'}^0_{L,R}$ and electroweak gauge bosons. 
Therefore, the dominant dark matter annihilation processes are
\bit
	\item $B' B' \to W^+W^-/ZZ$ via $t$-channel exchange of $H_L^\pm (H_L^0)$;
	\item $B' B' \to t\bar{t}$ via $t$-channel exchange of $T'$.
\eit
Since we have assumed the $H_L^\pm (H_L^0)$ are heavy, 
the $B' B' \to W^+W^-/ZZ$ process is suppressed compared to the $B' B' \to t\bar{t}$ process.
Thus it is very similar to the top flavored dark matter~\cite{Kilic:2015vka} with dark matter being vector boson. 
Unlike the fermionic top flavored dark matter, the $t$-channel process $B' B' \to t\bar{t}$ 
provides us the $s$-wave component of the dark matter annihilation without chirality suppressed. 
Approximately, the $s$-wave $B' B' \to t\bar{t}$ annihilation cross section is written as
\bea  
(\sigma v)_{B' B' \to t\bar{t}} \simeq  \frac{2 N_c g'^4 Y^2}{9 \pi } \frac{m_{B'}^2}{(m_{B'}^2 + m_{T'}^2)^2},
\label{eq:relict}
\eea
where $N_c$ is the color factor and $Y$ is the top quark charge under $B'$.
Having known the value of $g'$, we could determine the relation between  between $m_{B'} = \frac12 g' f$ and $m_{T'} = M$ from 
the dark matter relic abundance measurements.
If only the $B' B' \to t\bar{t}$  channel is dominant,  the thermal relic abundance $\Omega_{\rm dm}h^2 \simeq 0.12$ puts constraint on the parameters $(m_{B'}, m_{T'})$, 
as shown in Fig.~\ref{fig:ddsd}. 

\begin{figure}[htp]
\begin{center}
\includegraphics[width=0.17\textwidth]{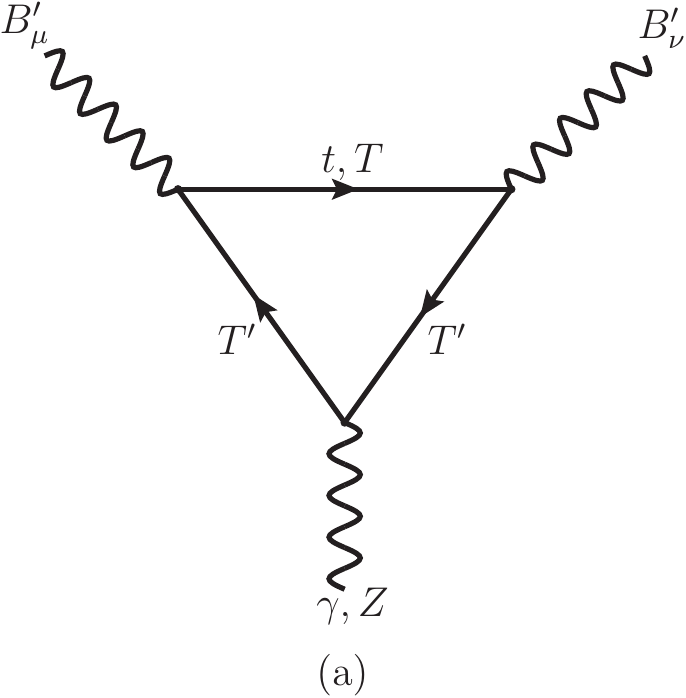} 
\includegraphics[width=0.17\textwidth]{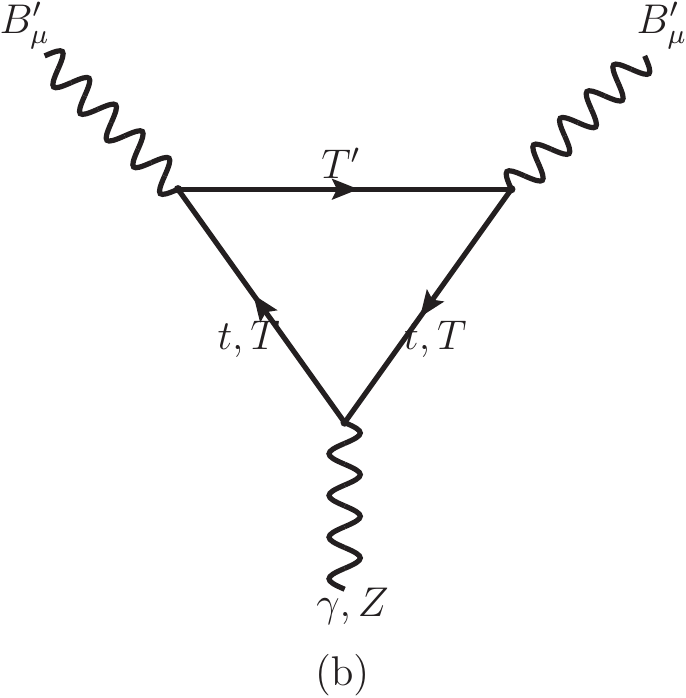} \\
\includegraphics[width=0.15\textwidth]{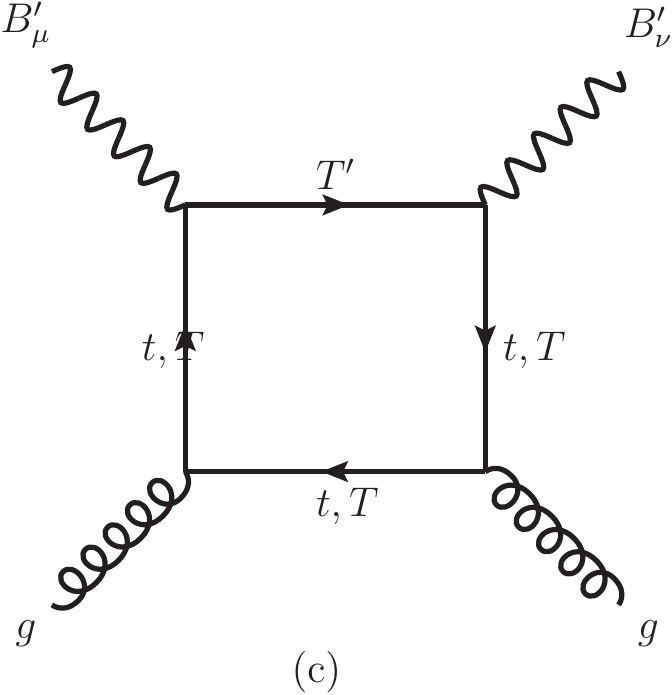} 
\includegraphics[width=0.15\textwidth]{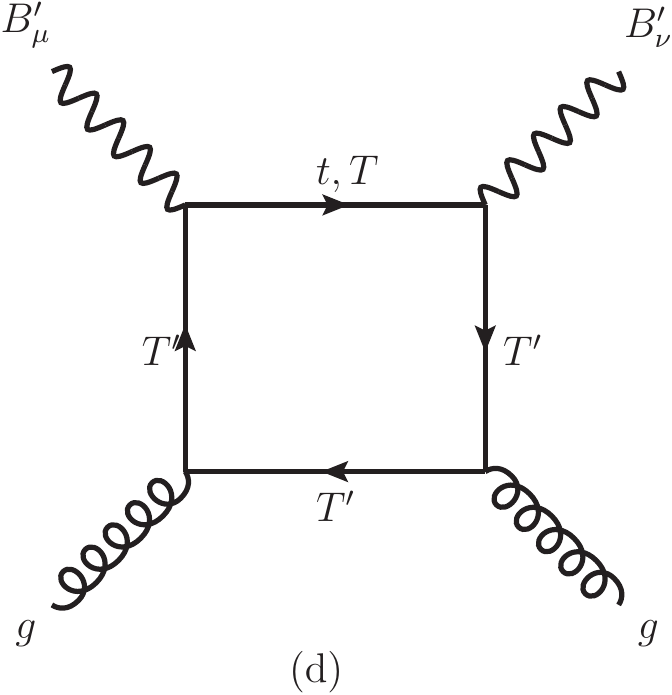} 
\includegraphics[width=0.15\textwidth]{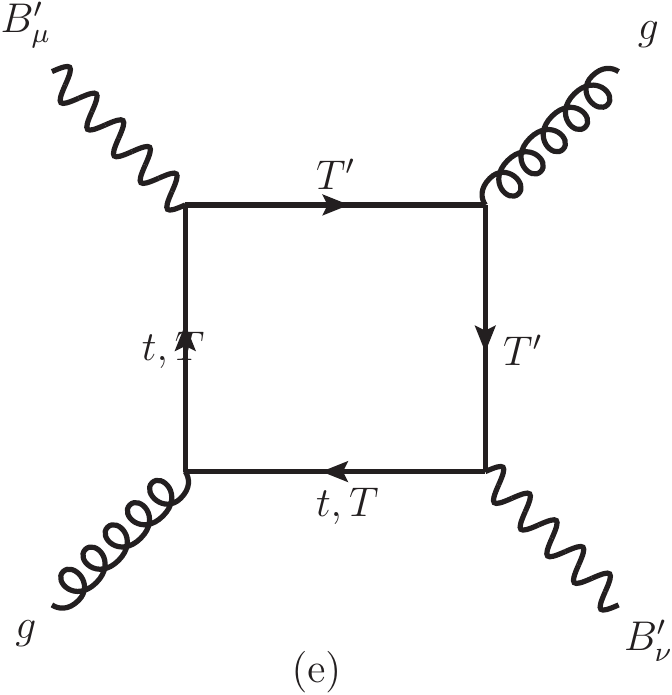} 
\caption{\small  The triangle (a,b) and box (c,d,e) Feynman diagrams which contribute
to the dark matter direct detection. }
\label{fig:feyndd}
\end{center}
\end{figure}

\begin{figure}[htp]
\begin{center}
\includegraphics[width=0.4\textwidth]{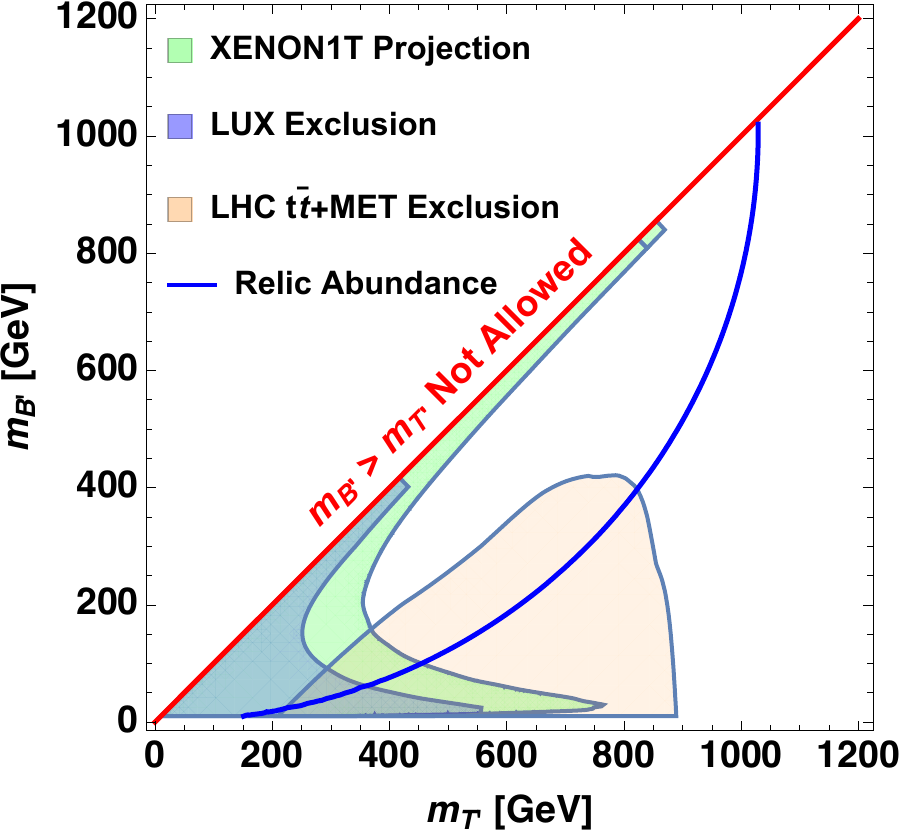} 
\caption{\small  The excluded contours on the model parameters $(m_{T'} = M, m_{B'} = \frac12 g' f)$ by the LUX experiments, XENON1T projection, and the top pair plus transverse missing energy searches at the LHC. The blue line shows the prediction of the dark matter relic abundance if only the $t$-channel $B'B' \to t\bar{t}$ is donimant. }
\label{fig:ddsd}
\end{center}
\end{figure}

We expect that the dark matter direct detection experiments will put further constraints on the model parameters. 
Because the $B'$ does not couple to the Higgs boson, the dominant contributions to 
direct detection come from the one-loop triangle and box diagrams, similar to the studies in Ref.~\cite{Yu:2014pra}. 
Let us calculate the low energy effective Lagrangian. 
According to Feynman diagrams shown in Fig.~\ref{fig:feyndd}(a,b), the resulting effective Lagrangian is
\bea
	{\mathcal L} =  {\mathcal C}_{\rm tri} \epsilon^{\mu\nu\rho\sigma} \left({B'}_\mu\partial_\nu {B'}_\rho\right) \overline{q}\gamma_\sigma \gamma_5 q\,,
	\label{eq:sdtriloop}
\eea
where the Wilson coefficient ${\mathcal C}_{\rm tri}$ is
\bea
	{\mathcal C}_{\rm tri}  &&\simeq N_c \frac{eQ g'^2}{\pi^2} \frac16 \int_0^1  {\rm d} z \frac{z^3 }{m_t^2 + (m_{T'}^2 - m_t^2)z + m_{B'}^2 z(z-1)} \nn\\
	&& + (m_t \leftrightarrow m_{T'}).
\eea
The box diagrams shown in Fig.~\ref{fig:feyndd}(c,d,e) generate the following effective Lagrangian
\bea
	{\mathcal L} = \alpha_s  {\mathcal C}_{\rm box}   {B'}^{\rho} {B'}_{\rho}G^{a\mu\nu}G^a_{\mu\nu},
	\label{eq:boxLag}
\eea
where the Wilson coefficient ${\mathcal C}_{\rm box}$ is approximately
\bea
	{\mathcal C}_{\rm box} =  \frac{g'^2}{48\pi} \frac{3 m_{T'}^2- 2m_{B'}^2}{(m_{T'}^2-m_{B'}^2)^2}.
\eea
The triangle loop diagrams only contribute to the spin-dependent (SD) cross section, while the box loop diagrams
contribute to the spin-independent (SI) cross section:
\bea
	\sigma^{\rm SI}_N &=&  \frac{\mu_N^2}{\pi} \left(\frac{4}{9}f^{(N)}_{TG} \frac{m_N}{m_{B'}} {\mathcal C}_{\rm tri} \right)^2, \nn\\
	\sigma^{\rm SD}_N &=&  \frac{16\mu_N^2}{\pi} \left( e  \sum_{q} \Delta^N_q {\mathcal C}_{\rm box}\right)^2,
\eea
where  $f^{(N)}_{TG}$ and $\Delta^N_q$ are defined in the Appendix of Ref.~\cite{Yu:2014pra}.
Then we utilize the complete explosure of the LUX 2016 data~\cite{Akerib:2016vxi} 
to constrain the dark matter spin-independent (SI) cross section. 
We found that the exclusion limit is quite weak, as shown in Fig.~\ref{fig:ddsd}.
Then we also use the projected exclusion reach of the XENON1T experiment with an exposure of 2.2 ton years~\cite{Aprile:2012zx}
to see whether the parameter region could be excluded by future experiments.
Fig.~\ref{fig:ddsd} shows that the dark matter direct detection could not impose strong constraints on the model parameters.

On the other hand, the collider searches put stronger limits on these $\mathbb{T}$-odd particles.
The $\mathbb{T}$-odd $T'$ should have large production cross section because of the QCD production mechanism,
and then subsequently decay to the top quark and the dark matter $B'$: 
\bea
p p \to T' \overline{T'} \to t B' \bar{t} B',
\eea
which appears as the top quark pair plus transverse missing energy final states. 
This final states have been investigated at the LHC by both ATLAS~\cite{Aad:2015pfx} and CMS~\cite{Khachatryan:2016oia}. 
Although the LHC searches on this final states focus on the exclusion on the stop quark, 
it shares the same event topology as the $\mathbb{T}$-odd $T'$ searches. 
Thus we could ultilize the LHC exclusion limits on the stop quark,
and recast the results of the exclusion limits into the exclusion limit of the $\mathbb{T}$-odd $T'$  mass. 
To recast the exclusion limit, we assume the same cut efficience in these processes and perform 
a simple scaling of the NNLO cross section of the stop quark to the NNLO cross section~\cite{Aliev:2010zk, GoncalvesNetto:2012yt}. 
We use the ATLAS analysis on stop quark searches with 20.3 fb$^{−1}$ integrated luminosity at 8 TeV~\cite{Aad:2015pfx},
and obtain the exclusion limit on $(m_{T'}, m_{B'})$, as shown in Fig.~\ref{fig:ddsd}.  
%
%
From Fig.~\ref{fig:ddsd}, we note that exclusion limit from the collider searches 
put much tighter constraints on the model parameters than the direct detection constraints. 
Combined all the constraints from Fig.~\ref{fig:ddsd} and Fig.~\ref{fig:higgs} together, we find that the mass parameter needs to be greater $900$ GeV by the collider searches and at the same time the scale $f$ needs to be greater than $1.5$ TeV via the Higgs coupling measurements.

Finally, the collider searches should also provide constraints on the ${\mathbb{T}}$-odd 
$H'$. Since $m_{H'} > m_{B'} \sim 400 $ GeV,
we expect that mass of the $H'$ should be heavier than 400 GeV. 
We know that typically the electroweak production limit is around 400 GeV. 
So compared to the limits from the $T'$ searches and the Higgs coupling measurements, the 
current exclusion limit on the $m_{H'}$ searches will not provide additional constraints. 
%
But we expect that the future searches on $H'$ might also put strong constraints
on the model parameters.


\section{Summary and Discussion}
\label{sec:conclude}

We have investigated implementing the $\mathbb{T}$-parity in the twin Higgs scenarios. 
This provides us $\mathbb{T}$-odd hidden sector and a promising dark matter candidate. 
We focused on one specific realization of this new scenario: 
the $\mathbb{T}$-parity extension of the left-right twin Higgs model. 
And we discussed collider constraints on the $\mathbb{T}$-even particles, and
dark matter phenomenology of the  $\mathbb{T}$-odd sector. 
We found that the tightest constraints come from the combination of the 
Higgs coupling measurements and the $\mathbb{T}$-odd top partner searches at the LHC.

This $\mathbb{T}$ parity twin Higgs model could be generalized to construct
other twin Higgs models.
And the interplay between the $\mathbb{T}$ parity and the $\mathbb{Z}_2$ symmetry
might generate new ideas on the twin Higgs models. 
For example, we might be able to construct a theory in which
the  $\mathbb{Z}_2$ symmetry only effects in the Higgs sector, but the $\mathbb{T}$-parity 
doubles the mass spectra. 
Overall, the implementation of the $\mathbb{T}$ parity in the twin Higgs model might
provide a new approach to understand the twin Higgs scenarios.


\begin{acknowledgments}
This work was supported by DOE Grant DE-SC0011095. 
\end{acknowledgments}

\appendix


\begin{widetext}

\section*{Mass Matrix Diagonalization}


Let us calculate the gauge boson masses. 
The $\mathbb{T}$-even charged gauge boson masses are 
\bea
	m_{W}^2 = \frac{1}{2} g^2 f^2 \sin^2 x, m_{W'}^2 = \frac{1}{2} g^2 f^2 \cos^2 x.
\eea
The $\mathbb{T}$-even neutral gauge bosons are mixed togther. 
Their mass matrix is written as
\bea
\left(\begin{array}{c|ccc}&W_L^3&W_R^3&B\\\hline
W_L^3&\frac{1}{2}g^2 f^2 \sin^2 x & 0 &-\frac{1}{2}gg' f^2 \sin^2 x\\
W_R^3& 0 &\frac{1}{2}g^2 f^2 \cos^2 x &-\frac{1}{2}g g' f^2 \cos^2 x \\
B&-\frac{1}{2}g g' f^2 \sin^2 x &-\frac{1}{2}g g' f^2 \cos^2 x &\frac{1}{2} {g'}^2 f^2\end{array}
\right).
\eea
Let us define the coupling constants
\bea
	g = \frac{e}{s_w}, g' = \frac{e}{\sqrt{c_{w}^2-s_{w}^2}} = \frac{e}{c_{2w}}, g_Y = \frac{e}{c_w},
\eea
with the weak mixing angle defined as $s_w = \sin\theta_w$ and $c_w = \cos\theta_w$.
It is more convenient to work in the basis $(A,Z_L,Z_R)$, where
\bea
\left(\begin{array}{c}A\\Z_L\\Z_R\end{array}\right)=
\left(\begin{array}{ccc}
s_{w}&s_{w}&\sqrt{c_{2w}}\\
- c_{w}&  s_{w}t_{w} & t_{w}\sqrt{c_{2w}}\\
0& -\frac{\sqrt{c_{2w}}}{c_w}& t_{w}
\end{array} \right)
\left(\begin{array}{c}W_L^3\\W_R^3\\B\end{array}\right). 
\eea
In this basis, the mass eigenvalue of the gauge boson $A$ is identically zero.
We identify this gauge boson $A$ is the photon that should remain massless after symmetry breaking. 
In this basis, the mass matrix reduces to
\bea
\left(\begin{array}{c|ccc}&A &Z_L& Z_R\\\hline
A  & 0 & 0 &0 \\
Z_L& 0 &M_{LL}^2 & M_{LR}^2\\
Z_R& 0 &M_{LR}^2 & M_{RR}^2 
\end{array}
\right),
\quad {\rm with} \quad
\begin{array}{l}
	M_{LL}^2 = \frac{g^2 + g_Y^2}{2} f^2 \sin^2 x,\\
	M_{LR}^2 = \frac{g_Y\sqrt{2{g'}^2 - g_Y^2}}{2} f^2 \sin^2 x,\\
	M_{RR}^2 = \frac{g^2 + {g'}^2}{2} f^2 - \frac{g^2 + g_Y^2}{2} f^2 \sin^2 x.
\end{array}
\eea
The exact eigenstates $Z,Z'$ are obtained via the rotation
\bea
\left(\begin{array}{c}Z\\Z'\end{array}\right)=\left(\begin{array}{cc}
\cos\vartheta& \sin\vartheta\\
-\sin\vartheta&\cos\vartheta\end{array} \right)
\left(\begin{array}{c}Z_L\\Z_R\end{array}\right), 
\quad \tan2\vartheta=\frac{2M_{LR}^2}{M_{LL}^2-M_{RR}^2}.
\eea
The eigenvalues $M_Z^2$ and $M_{Z'}^2$ are given by
\bea
M_{Z,Z'}^2=\frac{1}{2}\left(M_{LL}^2+M_{RR}^2\mp\left(M_{RR}^2-M_{LL}^2\right)\sqrt{1+\tan^22\vartheta}\right).
\eea
and approximately we have
\bea
	M_Z^2 = M_{LL}^2 - \frac{M_{LR}^4}{M_{RR}^2 - M_{LL}^2}, \quad M_{Z'}^2 = M_{RR}^2 + \frac{M_{LR}^4}{M_{RR}^2 - M_{LL}^2}.
\eea

For the top quark sector, the Yukawa terms can be rewritten as 
\bea
-{\mathcal L}_{\rm top} = \left(\begin{array}{c} t_L \\ T_L \end{array}\right) \left(\begin{array}{cc}
0 & y H_R\\
y H_L & M \end{array} \right)
\left(\begin{array}{c}t_R\\T_R\end{array}\right).
\eea
This  gives rise to the following mass matrix squared
\bea
{\cal M}^2_{\rm top}   = \left(\begin{array}{cc}
y^2 H_L^\dagger H_L & y M H_L\\
y M H_L & M^2 + y^2 H_R^\dagger H_R \end{array} \right).
\eea
Defining the mass eigenstates
\bea
\left(\begin{array}{c} {t}^{\rm mass}_{L,R} \\ {T}^{\rm mass}_{L,R} \end{array}\right) = \left(\begin{array}{cc}
\cos\alpha_{L,R} & \sin\alpha_{L,R}\\
-\sin\alpha_{L,R} & \cos\alpha_{L,R} \end{array} \right)\left(\begin{array}{c} t_{L,R}  \\ T_{L,R} \end{array}\right),
\tan2\alpha_{L,R} = \frac{4 y M H_{R,L}}{M^2 + y^2 H_{L,R}^2 - y^2 H_{R,L}^2},
\eea
we could diagonalize the mass matrix squared, and 
obtain the mass eigenstates
\bea
	m_{t,T}^2 &=&\frac12 (M^2 + y^2 f^2 \mp \sqrt{(M^2 + y^2 f^2)^2 - y^4 f^4 \sin^2 2 x }).
\eea
The mixing angle is
\bea
	\sin\alpha_L &=& \sqrt{1- (y^2 f^2 \cos 2x + M^2)/\sqrt{(M^2 + y^2 f^2)^2 - y^4 f^4 \sin^2 2 x }},\\
	\sin\alpha_R &=& \sqrt{1- (y^2 f^2 \cos 2x - M^2)/\sqrt{(M^2 + y^2 f^2)^2 - y^4 f^4 \sin^2 2 x }}.
\eea

\end{widetext}



\end{document}